\documentclass[runningheads,a4paper]{llncs}

\usepackage{amssymb}
\usepackage{graphicx}

\usepackage{url}

\newcommand{\keywords}[1]{\par\addvspace\baselineskip
\noindent\keywordname\enspace\ignorespaces#1}

\begin{document}

\mainmatter  

\title{Teleportation of a quantum state of a spatial mode  with a single massive particle}

\titlerunning{Mode teleportation with a single massive particle}

%
%
\author{Libby Heaney}

\authorrunning{Mode teleportation with a single massive particle}

\institute{Department of Physics, University of Oxford, Clarendon Laboratory, \\Oxford, OX1 3PU, UK
}

%
%

\toctitle{Teleportation of a quantum state of a spatial mode  with a single massive particle}
\tocauthor{Libby Heaney}
\maketitle

\begin{abstract}
Mode entanglement exists naturally between regions of space in ultra-cold atomic gases.  It has, however, been debated whether this type of entanglement is useful for quantum protocols.  This is due to a particle number superselection rule that restricts the operations that can be performed on the modes.  In this paper, we show how to exploit the mode entanglement of just a single particle for the teleportation of an unknown quantum state of a spatial mode.  We detail how to overcome the superselection rule to create any initial quantum state and how to perform Bell state analysis on two of the modes.  We show that two of the four Bell states can always be reliably distinguished, while the other two have to be grouped together due to an unsatisfied phase matching condition.  The teleportation of an unknown state of a quantum mode thus only succeeds half of the time.

\end{abstract}

\section{Introduction}

Entanglement is a key resource in many practical applications using quantum mechanics \cite{Nielsen:00}.  Usually entanglement is thought to exist between the degrees of freedom of two, or more, well localised quanta, such as photons or massive particles.   When particles are well separated, they are distinguishable from one another and can thus be assigned labels.   Entanglement between the particles is then well defined since their Hilbert space has a tensor product structure.   However, if the de Broglie wavelengths of identical particles begin to overlap, the particles become indistinguishable and one can no longer assign a label to each particle. The concept of standard particle entanglement breaks down as the Hilbert space no longer has the required tensor product structure; it is a projection onto the symmetric or antisymmetric subspaces (depending on the particle statistics) \cite{Zanardi:01}.  

While entanglement between indistinguishable particles can still be  correctly defined by taking a set of detectors into account \cite{Tichy:09}, another method to recover a tensor product Hilbert space is to move into the formalism of second quantisation.  Here, in the so called Fock basis, one defines a complete set of single particle states and counts the number of excitations in each.  The corresponding mode structure has a tensor product Hilbert space and hence entanglement of modes is a meaningful concept.  
Entanglement can therefore exist between modes occupied by particles \cite{Zanardi:01,Peres:95}.  The modes can be energy eigenmodes or perhaps more relevant to quantum communication or information processing protocols are spatial modes.  In the following section, we will discuss in more detail  the mode structure for a simple many-body system.  
The particles may be massless, i.e. photons, or massive, such as those found in ultra-cold gases.  
Mode entanglement of photons has been considered in a number of works \cite{Tan:91,vanEnk:06} and is usually limited to the single photon regime \cite{Hardy:94,Greenberger:95,Dunningham:07,Heaney:09}.  The experimental confirmation of the mode entanglement of a single photon via a Bell like test was obtained in 2004 \cite{Hessmo:04} and the multipartite entanglement of one photon distributed between four optical modes was characterized using uncertainty relations in an interferometric setup in 2009 \cite{Papp:09}.

For massive particles, the existence of entanglement between spatial modes becomes less clear \cite{Wiseman:03}.  This is due to a particle number superselection rule \cite{Wick:52,Giulini:96}, that forbids isolated systems from existing in a superposition of eigenstates of different mass.  Hence, for modes occupied by massive particles the system density operator, $\hat \rho$, cannot contain any off-diagonal terms that connect states of different particle number.   Any measurements made on the spatial modes are also restricted to the subspace of fixed particle number.  Since the correlations of entanglement are locally basis independent, they can be confirmed via for instance a Bell inequality, where one should measure each subsystem locally in at least two basis.  For spatial modes, the particle number is one such basis, however to measure in a second, rotated basis, measurements of superpositions of different numbers of particles are required.  For an isolated system, such superposition measurements are forbidden due to the number of particles in the system being fixed.  Until recently it was unresolved how to measure modes of a massive bosonic field in any  way other than in the particle number basis. And this is why it has been debated  \cite{Greenberger:96} whether mode entanglement is as `genuine' as particle entanglement or whether it is just a mathematical feature of the quantum state.  

Recent research has, however, shown that superselection rules \cite{Bartlett:07,Aharonov:67,Mirman:69} can be overcome locally by using a suitable reference frame \cite{Bartlett:07,Dowling:06}.  The reference frame required to rotate spatial modes away from the particle number basis is a coherent reservoir of particles such as a Bose-Einstein condensate (BEC).  By using such a reservoir, one can create, at least in principle,  superpositions of different numbers of particles \cite{Recati:05,Dowling:06}.  Thus, it is predicted that mode entanglement of a single massive particle is, indeed, as genuine as particle entanglement \cite{Cunha:07}.  Specific schemes for Bell inequality tests of mode entanglement of one  \cite{Ashhab:07,Ashhab:09,Heaney:09a}, and more \cite{Lee:InPrep}, massive particles have been given.  Another question is whether mode entanglement can be used for quantum communication and quantum information processing.  This is an interesting point, as mode entanglement occurs naturally in coherent ultra-cold bosonic gases \cite{Simon:02,Anders:06,Heaney:07,Libbythesis:08,Goold:09}.  If one could harness this entanglement for practical applications, it could by-pass the need to create complicated entangled states manually.  

A first step to understanding the usefulness of the mode entanglement of just a single massive particle came in a recent paper \cite{Heaney:09b}, which gave a scheme for implementing the quantum dense coding protocol.  Dense coding \cite{Bennett:92} allows the transmission of two classical bits of information via one qubit; in order to achieve the full quantum channel capacity a maximally entangled Bell state is initially required and full Bell state analysis is needed.  It was shown \cite{Heaney:09b} that the linear photonic dense coding channel capacity could be achieved without a BEC reservoir and that with a BEC reservoir the full quantum dense coding protocol could be implemented.  

While it was briefly mentioned in \cite{Heaney:09b} that the teleportation of a quantum state of a spatial mode should also be possible, no detailed scheme was given.  It is the aim of this paper, to provide such a scheme.  Quantum teleportation \cite{Bennett:93} is a key protocol in quantum information science and has been fruitfully demonstrated in experiments with a number of different physical implementations \cite{Bouwmeester:97,Furusawa:98,Boschi:98}.  In \cite{Lombardi:02}, a qubit (or optical mode) consisting of the vacuum and one photon states was teleported using the entanglement of a single photon in a superposition of two spatial modes.  Even though the creation of a superposition of different numbers of photons is not strictly forbidden by a superselection rule as it is with massive particles, there is still the problem of how to keep track of the phase between the two different photon number states.  Lombardi {\it et al} \cite{Lombardi:02} solved this by teleporting a mode that was entangled to another one.  In other words, the single photon was coherently distributed across the mode to be teleported and an ancilla mode.  This ancilla optical mode actually played the role of a phase reference in a similar way to how the BEC reservoir will play the role of a phase reference in our scheme here.  

One motivation to study a teleportation scheme using the mode entanglement of a single massive particle is to allow for direct comparisons of quantum information processing with different physical systems.  Moreover, from a fundamental viewpoint, it is interesting to clearly demonstrate that mode entanglement of massive particles is, in principle, useful entanglement - and can actually be used in a very similar way to particle entanglement - for quantum information processing despite the particle number superselection rule.  
In particular, in this paper we will see that in contrast to the dense coding scheme of \cite{Heaney:09b}, where the full channel capacity was attainable, here an arbitrary state of a quantum mode is only reliably teleported half of the time.  
This is due to an additional phase locking criterion that arises from having a total of three modes in the teleportation scheme as opposed to the two modes that are required for dense coding.  This illustrates subtle intricacies of teleportation with mode entanglement of a massive particle that are not present in the particle entanglement case.    

We begin in the subsequent section, by reviewing the concept of mode entanglement.  In section (\ref{sec:tele1}), we detail how to teleport the unknown state of a spatial mode using a single massive particle distributed coherently across two spatial modes.    We begin next by introducing the concept of spatial mode entanglement in more detail.  

\section{Entanglement of spatial modes}

We will start by detailing the mode structure of a bosonic system and then use a simple example of mode entanglement to illustrate the differences between it and particle entanglement.   Finally, we will explain some results concerning mode entanglement of Bose gases at zero and finite temperatures.   Note that we will consider in this paper only mode entanglement of bosonic fields; for a discussion of mode entanglement in fermionic systems see, for example, this paper  \cite{Aharonov:00} by Aharonov and Vaidman.

Consider  a confining volume, $V$, whose energy eigenmodes, labelled by $k$, can be `excited' by applying creation operators, $\hat a^{\dag}_k$, on the vacuum state, $\hat a^{\dag}_k |vac \rangle = |0_1 0_2 ... 1_k 0_{k+1} ... \rangle$, where $[\hat{a}_k,\hat a^{\dag}_l]=\delta_{kl}$.   An excitation is `a particle' with corresponding energy $E_k = \hbar \omega_k$, where $\omega_k$ is the frequency of the $k$-th energy eigenmode.  As mentioned already, instead of describing the system using its energy modes, it is often desirable to use a different set of modes, such as the \emph{spatial modes}. The description of the system in space is obtained by a transformation via the energy eigenfunctions, $
 \hat a^{\dag}_k | vac \rangle = \int  dx \, \phi_k(x) \, \hat \psi^{\dag}(x) \,| vac \rangle$,
 where $\phi_k(x)$ is the $k$-th energy eigenfunction and $\hat \psi^{\dag}(x)$ creates a particle at point, $x$, in space.  It follows that $[\hat\psi(x),\hat\psi(x')]=\delta(x-x')$. Populating an energy mode with a particle is thus equivalent to populating all spatial modes, i.e. all points in space, in a superposed manner. 

To illustrate the differences between particle and mode entanglement,
we take two non-interacting bosons trapped in a confining volume at zero
temperature.  In first quantisation, i.e. in the language that one would use to describe particle entanglement, the wavefunction is the
symmetrized product, $\Psi_{12}(x,y) = \frac{1}{\sqrt{2}}
(\phi_1(x)\phi_2(y)+\phi_1(y)\phi_2(x))$, where $\phi(x)$ is the
ground state of the confining potential. No entanglement exists
between the particles, since indistinguishability forbids us from
assigning to any particle a specific set of degrees of freedom.  In other words, the state space of the two particles is a projection onto the symmetric subspace, whereas a tensor product Hilbert space, $\mathcal{H} = \mathcal{H}_1\otimes\mathcal{H}_2$, is required to define entanglement between the subsystems.  

Conversely, in second quantisation one can define a pair of spatial
modes, $A$ and $B$, where each mode occupies half the confining
geometry.  Since both the particles are {\it coherently} distributed over
these modes, the system is described by the entangled
state, 
\begin{eqnarray}
|\psi\rangle &=& \frac{(\hat a^{\dag}_0)^2}{2}|vac\rangle\nonumber\\
&=&\frac{1}{2}\left(\frac{\hat{\psi}_A^{\dag}+\hat{\psi}_B^{\dag}}{\sqrt{2}}\right)^2|vac\rangle\nonumber\\
&=&\frac{1}{2}(|20\rangle+\sqrt{2}|11\rangle+|02\rangle),
\end{eqnarray}
 where  $ \hat\psi_X^{\dag} = \int_X dx\, g(x) \hat\psi^{\dag} (x)$ creates a particle in mode $X = A,\, B$  ($g(x)$ is a so called detector profile that gives weighting to the points in space), $|mn\rangle=|m\rangle_A\otimes|n\rangle_B$ span the state space $\mathcal{H} = \mathcal{H}_A\otimes \mathcal{H}_B$ and  $m$ denotes the number of particles in mode $A$ and $n$ the number particles in mode $B$ (with $m+n=2$).
From this example it is clear that entanglement is contingent
on the choice of modes \cite{Vedral:03}, but provided that like here a suitable choice is made,
investigating entanglement between {\sl distinguishable} modes circumvents the difficulties of defining entanglement between indistinguishable particles \cite{Esteve:08}.

 Spatial mode entanglement of an non-interacting BEC at zero temperature was first considered by Simon \cite{Simon:02}.  He and others \cite{Toth:03} found that mode entanglement existed between regions of space if the gas had an uncertainty in  particle number below a given level.  That is, if the gas is best described by a coherent state, $|\alpha\rangle$ (or mixtures there of), there is no entanglement between spatial modes.  On the other hand, if a uniform gas is of a fixed particle number, $N$, there is $\frac{1}{2}\log_2 N$  amount of entanglement (as measured by the von Neumann entropy), between two equal sized modes. Anders {\it et al} \cite{Anders:06} used a thermodynamical entanglement witness to show that spatial mode entanglement only exists across an entire Bose gas below the critical temperature for BEC.  
Spatial entanglement between two and more modes of an interacting Bose gas at finite-temperature with a fixed (but possibly unknown) number of particles was considered by Goold {\it et al.} \cite{Goold:09}, who demonstrated that mode entanglement is present in a gas when the coherence length of the particles extends over the modes.  More specifically, there is a direct link between single-particle reduced density matrix \cite{Pitaevskii:03} (i.e. the visibility of interference fringes) between different regions of the gas and spatial mode entanglement.  This means that the natural mode entanglement of  a BEC has already been detected in experiments such as \cite{Bloch:00}, albeit indirectly.  More recently, mode entanglement generated by a single exciton was predicted to exist between different sites in the  photosynthetic FMO complex \cite{Sarovar:10}, (as a result of the experimentally verified quantum coherence in the molecule), demonstrating that this type of entanglement is relevant even in biological systems.

\section{Teleportation of a quantum state of a spatial mode using a single massive particle}\label{sec:tele1}

Since mode entanglement exists naturally within many systems, it is important to ask whether, at least in principle, this type of entanglement  is useful entanglement.  We will address the question by providing a scheme to show that a single massive particle that is coherently distributed over two spatial modes can, at least theoretically, be used to teleport the unknown quantum state of a spatial mode perfectly half of the time in spite of the superselection rule.

In the following, we will first introduce a Hamiltonian whose parameters can be  switched on and off to perform the gates.  
We will show that in order to overcome the superselection rule and to rotate the modes away from the particle number basis, a Bose-Einstein condensate should be used as a particle reservoir and also as a phase reference throughout the teleportation protocol.   We will end by introducing the quantum circuit that allows to perform the teleportation protocol.  

The standard qubit teleportation protocol \cite{Bennett:93} between two parties, $A$ for Alice and $B$ for Bob, can be split up into four parts: \\
(i) {\bf preparation stage}: the preparation of an unknown quantum state by a third party, Charlie, and also the distribution of an entangled Bell state between Alice and Bob, \\
(ii) {\bf Bell state measurement by Alice}: the prepared  qubit is passed to Alice who then makes a Bell state measurement on this and her portion of the entangled state and records the measurement outcome, \\
(iii) {\bf transmission of classical information}:  Alice sends Bob two bits of classical information that indicate which of the four Bell states was measured, and \\
(iv) {\bf single mode rotation by Bob}: Bob uses the classical information to select which operation to perform on his qubit leaving him with the original unknown quantum state.\\ 
We will now detail how to perform the four steps using mode entanglement of a single massive particle.
\\\\
The system consists of three spatial modes, $a$, $A$ and $B$.  Mode $a$ will be placed in an unknown state, which will be teleported to mode $B$.  Modes $A$ and $B$ will be in the maximally entangled state formed from a single particle.  The system is described by the Bose-Hubbard model with additional coupling of each mode to the BEC reservoir.  The Hamiltonian is written as follows
\begin{eqnarray}
\hat H_{BH} &=&-\frac{J_{AB}}{2}(\hat\psi_A^{\dag}\hat\psi_B+\hat\psi_B^{\dag}\hat\psi_A)-\frac{J_{aA}}{2}(\hat\psi_a^{\dag}\hat\psi_A+\hat\psi_A^{\dag}\hat\psi_a)\nonumber\\
&+& \sum_{i=a,A,B} U_i\, \hat n_i(\hat n_i - 1) + \sum_{i=a,A,B} E_i\,\hat n_i - \sum_{i=a,A,B} \frac{\Omega_i}{2}(\hat\psi^{\dag}_i\hat\psi_{res}+\hat\psi_{res}^{\dag}\hat\psi_i)
\end{eqnarray}
The first two terms represent the coupling between the modes, $A$ and $B$, and also modes, $a$ and $A$.  Modes $a$ and $B$ do not interact throughout the entire protocol.  The coupling between the modes can be turned on and off by varying the tunneling matrix element $J_{aA}$ ($J_{AB}$) by increasing/decreasing the height of the potential barriers between the wells by altering the intensity of the trapping laser.  

The parameter $U_i$ is the onsite interaction term, which is continuously set to be much larger than the other energy scale in the Hamiltonian.  This is achieved by ensuring that each potential is tightly confined in all three directions, as is the case for atomic quantum dots \cite{Recati:05}.  The resulting large nonlinear repulsive interaction between the particles means that only zero or one particles can exist in the mode at any instance.  In other words, by maintaining a high repulsivity between the particles the modes are forced to behave like qubits.

The next three terms in the Hamiltonian are the free energies $E_i$ of the individual modes.  The standard setting throughout the protocol will be $E_a=E_A=E_B$, but to change the phases of the modes relative to one another, a potential bias can be applied to a mode using, for instance, a dispersive laser pulse.  

The final set of terms in the Hamiltonian correspond to the coupling of each individual mode to the BEC reservoir.  We consider a Raman laser set up, which couples the two different trapping states of the system and reservoir atoms  \cite{Jaksch:04}.  The parameter, $\Omega=\int dx \phi_0(x)\Psi_0(x)\tilde{\Omega}$, is the effective Rabi frequency, where $\Psi_0(x)$ is the wavefunction of the BEC, $\phi_0(x)$ is the wavefunction of the atom in mode $a$ and $\tilde{\Omega}$ is the usual Rabi frequency.   

\subsection{Single-mode and two-mode gates}

We will now discuss the implementation of the single-mode and two-mode gates that will be used in the teleportation protocol.  

\subsubsection{Single-mode phase gate:}\label{sec:phase}

A phase gate, for instance a Pauli $\hat Z$ operation, can be applied to mode $j$ by altering the energy, $E_j$, relative to the free energies for the other modes for a given time, $t$.  The corresponding unitary operation on the mode is $\hat U = e^{i E_j \hat n_j t}$, which if applied for $t=\pi/E_i$ performs the $\hat Z$ gate, $|0\rangle\rightarrow |0\rangle, \, |1\rangle\rightarrow - |1\rangle$.  During this operation, the couplings between the modes themselves and between the modes and the BEC are switched off.

\subsubsection{Single-mode number rotation gate:}\label{sec:number-rotation}

We would also like to rotate the state of an individual mode between the $|0\rangle$ and $|1\rangle$ eigenstates.  To do this we need to couple to the BEC reservoir by applying the Raman pulse for a time, $t$. 
We take here a uniform non-interacting BEC described by a mixture of coherent states, $\hat\rho_{bec}=\int_0^{2\pi}\frac{d\theta}{2\pi}||\alpha|e^{i\theta}\rangle\langle|\alpha|e^{i\theta}|$, where $|\alpha|^2=\bar{n}$ is the average number of particles in the condensate and $\theta$ is the condensate phase.  
For a nummerical analysis of a single mode, i.e. an atomic quantum dot, coupled to a superfluid reservoir  of interacting particles see \cite{Lee:10}. 

Note that when calculating it suffices to use the coherent state, $||\alpha|e^{i\theta}\rangle$, with one realisation of the phase $\theta$  and to apply the twirling operator, $\mathcal{T}[\rho_L]=\int_0^{2\pi}\frac{d\theta}{2\pi}[\hat\rho_L]$, to obtain the quantum state of laboratory  as seen by Bob (here $\hat\rho_L$ is the density operator for the total system and laboratory \cite{Dowling:06,Bartlett:07}).  Note that, unlike the dense coding scheme \cite{Heaney:09b}, Alice and Bob do not necessarily need to use the same BEC for their operations in this teleportation scheme.
The unitary evolution governed by $\hat{H}_{int}= \frac{\Omega_i}{2}(\hat\psi^{\dag}_i\hat\psi_{res}+\hat\psi_{res}^{\dag}\hat\psi_i)$, transforms the occupation number of the modes as follows
\begin{eqnarray}
\label{Eq:modeBEC}
|0\rangle&\rightarrow&\cos(\frac{\Omega\sqrt{\bar{n}}t}{2})|0\rangle-i\sin(\frac{\Omega\sqrt{\bar{n}}t}{2})e^{i\theta}|1\rangle\nonumber\\
|1\rangle&\rightarrow&\cos(\frac{\Omega\sqrt{\bar{n}}t}{2})|1\rangle-i\sin(\frac{\Omega\sqrt{\bar{n}}t}{2})e^{-i\theta}|0\rangle.
\end{eqnarray}
Here we have traced out the BEC, which remains separable to the state of the modes as we assume it to have a high mean number of particles, $\bar{n}>>1$.  By coupling the modes to the BEC for various times, $t$, single mode number rotation gates can be performed.

\subsubsection{Two-mode c-phase type gate:}\label{sec:two-mode-gate}

We also need a two mode entangling gate to allow for Bell state analysis. A c-phase type  gate between two of the modes, $j\neq k \,\in a, A, B$, is performed as follows.  Since we consider a large onsite interaction strength throughout this protocol, the bosons behave like spin-polarised fermions.  One can see this by associating the two particle number states, $|0\rangle$ and $|1\rangle$, of each mode with the up/down spin half degree of freedom.  The bosonic Hamiltonian, $\hat H = -\frac{J}{2}(\hat\psi_j^{\dag}\hat\psi_k+\hat{\psi}^{\dag}_k\hat{\psi}_j)$, becomes equivalent to the quantum XX spin model, which in turn can be mapped via the Jordan-Wigner transformation to $\hat H=-J(\hat c_j^{\dag}\hat c_k+\hat c_k^{\dag}\hat c_j)$, where $\hat{c}_X^{\dag}$ and $\hat c_X$ are Fermionic creation and annihilation operators for mode, $X=j,\,k$ that anti-commute, $\{\hat c_X,\hat c_Y\}_+=\delta_{XY}$ \cite{Sachdev:00}.  Once the system is in this regime, the barrier between the modes, $j$ and $k$, is lowered for time, $t=\pi/(2J)$, so that the particles exchange position.  This ensures that the $|11\rangle$ term picks up a minus sign $|11\rangle\rightarrow-|11\rangle$, due to the anti-commutation relations \cite{Clark:05}.  The other three states transform as, $|00\rangle\rightarrow|00\rangle$ $|10\rangle\rightarrow|01\rangle$ and $|01\rangle\rightarrow|10\rangle$.  

\subsection{Circuit diagram for teleportation of a quantum state of a mode}

\begin{figure}[htbp] 
   \centering
   \includegraphics[width=4.5in]{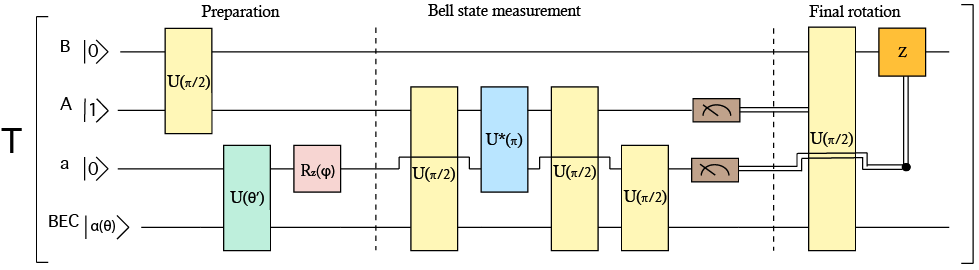} 
   \caption{The circuit for teleportation of a quantum state of a spatial mode.  The three modes and the BEC are represented in the figure by the four horizontal lines in the diagram and the gates are represented as boxes.  The first portion of the circuit is the {\it  preparation }of the entanglement via a single particle between Alice and Bob and the creation of an unknown state of the spatial mode $a$.  The {\it Bell state measurement} on modes $a$ and $A$ by Alice is outlined in the centre portion of the circuit.  The last part of the circuit details the {\it final rotation} on mode $B$ by Bob.  Further details will be discussed in the text. }
   \label{fig:teleportation_circuit}
\end{figure}

We now present the circuit diagram for the teleportation protocol, see Fig.  (\ref{fig:teleportation_circuit}), referring to the four steps outlined at the start of this section.  The three modes and the BEC are represented in the figure by the four horizontal lines in the diagram and the gates are represented as boxes.    
\\\\
{\bf The preparation stage:} a third party, Charlie, prepares a spatial mode ($a$) in an unknown quantum state by performing a single-mode number rotation (green in the diagram [section (\ref{sec:number-rotation})]) and a single-mode phase gate (red in the diagram [section (\ref{sec:phase})]) for the chosen amounts of time.  Initially, mode $a$, is in the vacuum state $|a\rangle=|0\rangle_a$, so that the two gates rotate it to the state $|a\rangle=\alpha|0\rangle+\beta e^{i(\theta+\phi)}|1\rangle$, where $\alpha = \cos(\theta'),\, \beta=-i\sin(\theta')$ and $\theta' = \Omega\sqrt{\bar{n}}t/2$ and $\phi=E_it$.  The potential barrier between modes $a$ and $A$ remains high throughout this stage.
An entangled Bell state shared between Alice and Bob is also required for the protocol.  To obtain such a state a single massive particle needs to be coherently distributed over the two modes -- if the single neutral atom initially starts in mode $A$, lowering the barrier to mode $B$ for time $t= \pi/(2J)$ creates the state $|\psi^+\rangle_{AB}=\frac{1}{\sqrt{2}}(|10\rangle+|01\rangle)$.  
\\\\
{\bf The Bell state measurement:} Spatial mode $a$ is now passed to Alice, who performs Bell state analysis on modes, $a$ and $A$.  Bob is in possession of mode $B$.      A measurement in the Bell state basis proceeds as follows:

(1)  Modes $a$ and $A$ should be within close proximity of one another and the potential barrier between them is high so that tunneling is fully suppressed.  Mode $A$ is rotated away from the particle number basis by coupling to a BEC reservoir via the Raman laser set-up for time $t=\pi/(2\Omega\sqrt{\bar{n}})$ as denoted by the yellow gate in the circuit diagram [see section (\ref{sec:number-rotation})].  Note that the BEC here does not necessarily need to be the same as the one used  by Charlie to create the initial state of mode $a$.  This results in the transformations, $|0\rangle_A\rightarrow\frac{1}{\sqrt{2}}(|0\rangle-ie^{i\theta}|1\rangle)_A$ and $|1\rangle_A\rightarrow\frac{1}{\sqrt{2}}(|1\rangle-ie^{-i\theta}|0\rangle)_A$.  For instance, the $|\psi_+\rangle_{aA}$ Bell state transforms (without normalisation) to $|01\rangle_{aA}+|10\rangle_{aA}\rightarrow|0\rangle_a(|1\rangle-ie^{-i\theta}|0\rangle)_A+|1\rangle_a(|0\rangle-ie^{i\theta}|1\rangle)_A$.  The other Bell states transform in a similar way.

(2) Then, a c-phase type  gate [see section (\ref{sec:two-mode-gate})] is applied by lowering the potential barrier between modes $a$ and $A$ for time $t=\pi/(2J)$ so that the bosons swap positions.  This is illustrated by the pale blue gate in the circuit diagram. 
By continuing the example in the previous step, one can see that this transformation works in the following way: 
$ |0\rangle_a(|1\rangle-ie^{-i\theta}|0\rangle)_A+|1\rangle_a(|0\rangle-ie^{i\theta}|1\rangle)_A\rightarrow |01\rangle_{aA}+|10\rangle_{aA} -ie^{-i\theta}|00\rangle_{aA}+ie^{i\theta}|11\rangle_{aA}
=(|0\rangle+ie^{i\theta}|1\rangle)_a(|1\rangle-ie^{-i\theta}|0\rangle)_A$.

(3) In order to rotate modes, $a$ and $A$, to the particle number basis for the read-out, each should be coupled to the BEC reservoir for time, $t= \pi/(2\Omega\sqrt{\bar{n}})$, as in step (2).    For example,  the state, $(|0\rangle+ie^{i\theta}|1\rangle)_a(|1\rangle-ie^{-i\theta}|0\rangle)_A$, after c-phase type gate above becomes $|00\rangle_{aA}$.  Hence a measurement of zero particles in both modes means that the quantum state, $|a\rangle=\alpha|0\rangle +\beta e^{i(\theta+\phi)}|1\rangle$ has been teleported to Bob.  A measurement of $|01\rangle_{aA}$ corresponds to the state, $\hat Z|a\rangle=\alpha|0\rangle -\beta e^{i(\theta+\phi)}|1\rangle$, being teleported to Bob's mode.   Note that the teleported states both also have phases which are correlated to the BEC, which means if they are to be used for other purposes then the {\it same} BEC would be needed to ensure phase matched conditions.

However, if there is one particle in mode $a$ after this final rotation, the teleportation has not succeeded.  This is due to the fact that after steps (1)-(3), the  states, $|\phi^{\pm}\rangle_{aA}$, are rotated to $|1\rangle_a((1\pm e^{i2\theta})|0\rangle+(1\mp e^{i2\theta})|1\rangle)_A/2$.   Since one does not know the phase, $\theta$ of the condensate, one has to average over all realisations of it, i.e. apply the twirling operator to determine the quantum state of the laboratory, which accounts for our ignorance.  This leaves mode $A$ in the maximally mixed state so that it is impossible to say which of the two operators, $\hat X$ and $\hat Z\hat X$, should be applied to recover the original state, $|a\rangle$, even in principle.  
\\\\
{\bf Transmission of classical information and final single-mode rotation}:   Alice thus sends one of three messages to Bob according to the outcome of her measurement.  Bob obtains the unknown quantum state half of the time upon the following transformations:
\begin{eqnarray}
|\psi^+\rangle_{aA}&\rightarrow&|00\rangle_{aA}\rightarrow \hat I_B,\quad\quad
|\psi^-\rangle_{aA}\rightarrow|01\rangle_{aA}\rightarrow \hat Z_B\nonumber\\
|\phi^{\pm}\rangle_{aA}&\rightarrow&|1\rangle\langle 1|_a\otimes\hat I_A \quad\rightarrow\quad \textrm{teleportation failed}
\end{eqnarray}
where the first column corresponds to the Bell state Alice measured, the second column to the particle number measurements in mode $a$ and $A$ (and also the classical bits that she sends) and the third column to Bob's operation on mode $B$.
The $\hat Z$ operation is applied to mode $B$ by changing the phase of the mode [as in section (\ref{sec:phase})] for time $t=\pi/E_B$ where $E_B$ is the energy bias of mode $B$.
This concludes the teleportation protocol for a unknown quantum state of a spatial mode.  

 \section{Discussion and conclusion}
 
 In this paper, we have discussed the teleportation of an unknown state of a spatial mode using a mode entangled state formed from a single particle.  We modeled the modes as qubits by considering tightly confined potentials, with zero or one particles representing the qubit degrees of freedom.  We have given an explicit scheme for creating an arbitrary state of a spatial mode despite the particle number superselection rule, by coupling to a BEC reservoir.  We have shown that one can reliably distinguish two of the four Bell states, with the other two grouped together.  This is due to the fact that we do not know the phase of the BEC reservoir, which is still classically correlated to the state after the Bell state analysis.
 However, the teleportation of a general unknown (two-level) state of a spatial mode is still achieved half of the time, which would never be possible if we could not locally bypass the superselection rule in the first place.  
    
This contrasts with a dense coding scheme using the mode entanglement of a single particle (see \cite{Heaney:09b}).  In this scheme, one can always discriminate all four Bell states.  The difference between the two schemes arises from the difference in symmetries of the encoding (or preparation) and decoding (or Bell state analysis) processes in the two protocols.  In the dense coding protocol, there are only two modes and Alice can initially couple her mode to the BEC to rotate to one of the two states, $|\phi^{\pm}\rangle=\frac{1}{\sqrt{2}}(|00\rangle\pm e^{i2\theta}|11\rangle$).  In doing so, the phase of these states becomes classically correlated to the phase of the BEC, $\theta$. Alice then passes her mode to Bob, who performs Bell state analysis as was outlined above to find which Bell state Alice sent.  Because the phase of the BEC is already present in $|\phi^{\pm}\rangle$ it cancels with the phase that is picked up again from the BEC during the Bell state measurement.  Thus all four Bell states are reliably distinguishable from one another.  

On the other hand, in the teleportation scheme, there are necessarily three modes.  Initially the mode that will be subsequently teleported to Bob is rotated to an arbitrary state and becomes classically correlated to the phase of the BEC -- {\it this phase will stay present in the state of this mode as it is teleported to Bob and does not feature in the Bell state analysis}.   Bell state analysis is then performed on this and one mode of the entangled state.  However, the phases that are picked up from the BEC in the Bell state analysis do not have any phases  to cancel with, unlike in the case of the dense coding scheme.  
So while we have access to all of the gates that are required to perform the teleportation protocol, because we cannot know the phase of the BEC, even in principle, we do not teleport the spatial mode all of the time.  This paper has illustrated how mode entanglement of massive particles does not behave exactly like particle entanglement despite the fact that one can locally overcome the particle number superselection rule.  Future work should focus on extending this scheme to consider the teleportation of a general $d>2$ level state of a spatial mode using the natural entanglement in BECs.


\begin{thebibliography}{99}
\bibitem{PhaseImprint} 
	Note that controlling the phase of a BEC in this way is known as phase imprinting; see e.g. \cite{Denschlag:00},
\end{thebibliography}
\end{document}